\documentclass[epj,nopacs,11pt]{svjour}
\usepackage{graphics}
\begin{document}
\title{A cross-validation check in the covariance analysis of isospin sensitive  observables
from heavy ion collision}
\author{I. Bednarek
\thanks{e-mail: ilona.bednarek@us.edu.pl}
\and J. S{\l}adkowski
\thanks{e-mail: jan.sladkowski@us.edu.pl}
\and J. Syska
\thanks{e-mail: jacek.syska@us.edu.pl}
%
%
}                     
%
%
\institute{Institute of Physics, University of Silesia, 75 Pu{\l}ku Piechoty 1,
Pl 41-500 Chorz{\'o}w, Poland \\
}

\date{Received: date / Revised version: date}
%

\authorrunning{I. Bednarek et al.}
\titlerunning{The cross-validation check in the covariance analysis of isospin sensitive  observables in HIC}

\abstract{
This paper focuses on two problems.  The first  is related to the consistency checks of the  complete correlation coefficients  between two groups of variables obtained in \cite{covariance-analysis-of-symmetry-energy}.  The first group  collects parameters that describe properties of nuclear matter referring to the transport model involved.   The second one is a group  of  observables adopted in the heavy ion collision experiments.  The second problem concerns the method of determining the values of  pure correlations between some variables in heavy ion collision.
 The application of this method for the analysis of the correlations of the isospin sensitive variables is pointed out.
}

%

\maketitle
\label{Introduction}




\section{Introduction}
\label{Introduction}

Attempts at explaining physical phenomena related to such diverse objects as atomic nuclei and neutron stars depend critically on proper description and understanding of the equation of state  (EoS) of asymmetric nuclear matter.
Heavy ion collisions (HICs) offer the opportunity to explore this EoS under conditions, which  depending on the energy of the beam and the isotopic composition, make it possible to reproduce and study the nuclear matter for various ranges of density $\rho$ and neutron-proton asymmetry $\delta =(\rho_{n}-\rho_{p})/(\rho_n+\rho_p)$, where $\rho_{n}$ and $\rho_{p}$ are neutron and proton densities.
So far,  significant constraints have been obtained on the symmetric part of the nuclear matter EoS \cite{2016:Baldo}.
However, constraints on its isospin dependent part are subject to the severe uncertainty,  especially in the high density limit.
Researches on asymmetric nuclear matter focus on developing  theoretical and experimental methods to overcome difficulties associated with the fact that the symmetry energy being encoded in the isospin dependent part of the EoS is not an observable.
Hence the overriding problem is to determine for  given physical conditions optimal observables that would allow one to impose reliable constraints on parameters, which describe asymmetric nuclear matter.
In particular the estimators of  parameters that enable to describe the symmetry energy dependence on density include, among others, the symmetry energy
 coefficient
$S_{0}$ and the slop of the symmetry energy $L$.
 As it will be shown, one
of the most important experiments that permit extraction of information about  correlations of the estimators of these parameters are HICs.
However, interpretations of data collected in measurements carried out in these experiments are not straightforward.
They are affected by various types of uncertainties in  the transport model involved. \\
This paper bases on the results reported in \cite{covariance-analysis-of-symmetry-energy}.
The covariance analysis presented there permited  to estimate the complete  linear correlation coefficients between
the variables, which are called
the force parameters and the group of observables adopted in the
HIC experiments to extract information on the nuclear matter EoS.  The force parameters are adopted in the transport model (i.e., the ImQMD-Sky model \cite{covariance-analysis-of-symmetry-energy}) wildly used in the interpretation of HIC data.
In the present, paper the formal relation between complete correlation coefficients obtained in  \cite{covariance-analysis-of-symmetry-energy} and partial correlation
coefficients \cite{Kleinbaum,Kendall} for the force parameters
is used (Eq.(\ref{partial correlation coeff})) to perform
an out-of-sample check for the mentioned complete correlation coefficients.
This  is a kind of the cross-validation check
for assessing if the complete correlation coefficients, obtained
from the one set of
HIC data, are consistent with
the complete correlation coefficients
obtained from another independent data set (see Section II).
\\
It was shown \cite{2005:Agrawal,2009:Chen}  that in the ImQMD-Sky transport model,  involving the effective Skyrme interactions, as the input variables the  parameters $(\rho_{0}, E_{0}, K_{0}, S_{0}, L, m_{s}^{\star}, m_{v}^{\star})$ can be used.
These parameters that mean successively: $E_{0}$ - the binding energy of nuclear matter, $K_{0}$ its incompressibility, $S_{0} \equiv S({\rho_{0}}) $ - the symmetry energy coefficient, $L$ slope of the symmetry energy, $m_{s}^{\star}$ and $m_{v}^{\star}$ the isoscalar and isovector effective masses, determine properties of nuclear matter at saturation density $\rho_{0}$.
 Analysis of the selected Skyrme forces, on the basis of which EoSs of asymmetric nuclear matter were obtained, had been described in  \cite{2016:Baldo}.
After obtaining the EoS, it becomes possible to derive the formula for the symmetry energy  $S({\rho})$ as a function of density  $\rho$. \\
Referring these results to the parameters describing  nuclear matter, it becomes attainable to determine, among others,  $S({\rho_{0}})$, $L$ and $K_{0}$ for the saturation density $\rho_{0}$. Expressions defining $S(\rho)$, $L$ and $K$ at arbitrary value of $\rho$ depend on various  sets of coefficients and  therefore, one should  expect a different degree of correlations between these quantities. Some of them may be physical.
\\
An example that  compiles data on selected 142 Skyrme interactions meeting strictly defined physical constraints is given in \cite{2016:Baldo}. In this case, taking into account the constraint associated with the reproduction of the binding energy, very weak correlation between $S_{0}$ and $L$ was obtained.
This led to the conclusion that given the symmetry energy  dependence on  density, it could be possible to extract correlations between $L$ and $S(\rho_{0})$ only if symmetry energy sensitive constraints are imposed \cite{2016:Baldo}.

Thus, it is resonable to  present a different approch to study the existence of nontrivial correlations between parameters that describe properties of nuclear matter. The analysis presented underneath gives, along with the above mentioned
cross-validation check on the consistency of the results on the complete correlation coefficients obtained from various HIC experiments \cite{covariance-analysis-of-symmetry-energy},
a method for determining the  existence of correlations between
 the force parameters. In this respect  only
the partial  correlation coefficients  give the pure nature of these correlations
\cite{Kleinbaum,Kendall}.



\section{Basic formulas and consistency conditions}

\label{Basic and consistency}


Zhang et al. \cite{covariance-analysis-of-symmetry-energy} performed the covariance analysis between two classes of variables. Firstly, the variables
 $A_{1} \equiv K_{0}$, $A_{2} \equiv S_{0}$ and $A_{3} \equiv L$ form, among others, the group $A$ of the force parameters. Secondly, the observables: 1) single $n/p$ ratio $B_{1} \equiv  CI-R_{2}(n/p) = \frac{Y_{2}(n)}{Y_{2}(p)}$, 2) double $n/p$ ratio $B_{2} \equiv  CI-DR(n/p) = CI-R_{2}(n/p)/CI-R_{1}(n/p) = CI-R_{21}(n/n)/CI-R_{21}(p/p)$, 3) isoscaling ratios $B_{3} \equiv  CI-R_{21}(n/n) = \frac{Y_{2}(n)}{Y_{1}(n)}$, 4)~$B_{4} \equiv  CI-R_{21}(p/p) = \frac{Y_{2}(p)}{Y_{1}(p)}$ and 5) the isospin transport ratios $B_{5} \equiv  R_{diff}$ form the group $B$ of five  isospin sensitive observables. In the above expressions $CI$ denotes the coalescence invariant nucleon yield spectra. $Y_{i}(n)$ and $Y_{i}(p)$ are integrated $CI$ neutron and proton yields from reaction $i$.
In the analysis performed in  \cite{covariance-analysis-of-symmetry-energy} the nucleon yield observables were obtained for the
the reactions
${}^{124}Sn + {}^{124}Sn$  and ${}^{112}Sn + {}^{112}Sn$ for 50 MeV and 120 MeV per nucleon, $\left[MeV/u\right]$. Additionally, in the case of $R_{diff}$ the mixed reaction ${}^{124}Sn + {}^{112}Sn$ was considered.
Following this notation the
out-of-sample check
for the complete, partial and multiple linear correlation coefficients between the variables $K_{0}$, $S_{0}$ and $L$ is below presented. The analysis concerns experimental data, discussed in \cite{covariance-analysis-of-symmetry-energy}, for two essential energy values 50 MeV/u and 120 MeV/u.

The complete correlation coefficients $r_{A_i B_a}$ between the variables $A_{i}$, $i=1,2,3$, and $B_{a}$, $a=1,2,...,5$, were obtained in \cite{covariance-analysis-of-symmetry-energy}.
Below, on this basis only, the possible information on the complete, partial and multiple correlation coefficients between the variables $A_{i}$ and $A_{j}$, $i,j=1,2,3$, $i \neq j$, is extracted.
Every partial correlation coefficient between the variables $A_{i}$ and $A_{j}$, $i,j=1,2,3$, $i \neq j$,  with the variable $B_{a}$,  $a=1,2,..,5$, which is under control, is given as follows \cite{Kleinbaum,Kendall}:
\begin{eqnarray}
\label{partial correlation coeff}
r_{A_{i}A_{j}|B_{a}} = \frac{r_{A_{i}A_{j}} - r_{A_{i}B_{a}} r_{A_{j}B_{a}}}{\sqrt{1 - r_{A_{i}B_{a}}^{2}} \sqrt{1 - r_{A_{j}B_{a}}^{2}}} \; . \;\;\;
\end{eqnarray}
The influence of $B_{a}$ on the correlation of $A_{i}$ and $A_{j}$ is removed  in $r_{A_{i}A_{j}|B_{a}}$ by the regression adjustment of $A_{i}$ with respect to $B_{a}$ and separately by the regression adjustment of $A_{j}$ with respect to $B_{a}$ \cite{Kleinbaum,Kendall}.
\\
If one assume that the equalities:
\begin{eqnarray}
\label{centralny war}
r_{A_{i}A_{j}|B_{a}} = r_{A_{i}A_{j}|B_{b}} \;\;\; {\rm for} \;\; a \neq b, \;\; a,b=1,2,..,5 \;
\end{eqnarray}
hold
then it follows from Eq.(\ref{partial correlation coeff}) that the value of the complete correlation coefficients $r_{A_{i}A_{j}} = r_{A_{j}A_{i}}$, $i,j=1,2,3$, $i \neq j$, are equal to:
\begin{eqnarray}
\label{r XY}
\!\!\!\! \!\!\!\! \!\!\!\! \!\!\!\!  \!\!\!\! \!\!
r_{A_{i}A_{j}}
&=&
\left(
r_{A_{i} B_{a}} r_{A_{j} B_{a}} \sqrt{1 - r_{A_{i} B_{b}}^{2}} \sqrt{1 - r_{A_{j} B_{b}}^{2}}
\right.
\nonumber
\\
\!\!\!\! \!\!\!\! \!
&-&
\left.
r_{A_{i} B_{b}} r_{A_{j} B_{b}} \sqrt{1 - r_{A_{i} B_{a}}^{2}} \sqrt{1 - r_{A_{j} B_{a}}^{2}} \;
\right)
\nonumber
\\
\!\!\!\! \!\!\!\! \!\!\!\! \!\!\!\! \!\!\!\! \!\!
&\times&
\left(\sqrt{1 - r_{A_{i} B_{b}}^{2}} \sqrt{1 - r_{A_{j} B_{b}}^{2}}
\right.
\nonumber \\
&-&
\left.
\sqrt{1 - r_{A_{i} B_{a}}^{2}} \sqrt{1 - r_{A_{j} B_{a}}^{2}} \; \right)^{-1} \, .
\end{eqnarray}
Thus, three complete correlation coefficients $(r_{A_{i}A_{j}}) = (r_{SK}, r_{LK}, r_{SL})$ are obtained.\\

{\it The consistency conditions.} If the correlations coefficients $r_{A_{i}A_{j}}$ and $r_{A_{i}A_{j}|B_{a}}$ are calculated from one sample obtained in the experiment $"a"$ then by the theorems \cite{Kleinbaum,Kendall},
they have to lie in the range $\langle -1, 1\rangle$.
Yet, if $r_{A_{i}A_{j}}$ is obtained from $r_{A_{i}B_{a}}$ via the comparison $r_{{A_{i}A_{j}}|B_{a}} = r_{{A_{i}A_{j}}|B_{b}}$, $a \neq b$, i.e., from different samples $"a"$ and $"b"$, as in Eq.(\ref{r XY}), then the requirement
$r_{A_{i}A_{j}} \in \langle -1, 1\rangle$ may fail and some
inconsistency can appear.
Then
$r_{A_{i}A_{j}} \in \langle -1, 1\rangle$ stands only for the {\it consistency condition},
giving the out-of-sample check for
the results
obtained from the considered different collision experiments.
This may be called
the different samples results consistency problem. \\
The sources of the inconsistency can be diverse.
Firstly, the {\it statistical} one,  different samples results inconsistency can appear as the result of the fact that the correlations $r_{A_{i}B_{a}}$ prescribed
to the observables $B_{a}$, $a=1,2,..,5$, were  (via the physical model) calculated \cite{covariance-analysis-of-symmetry-energy} for
different, finite accuracy experiments $"a"$.
Secondly, the {\it theoretical} inconsistency can appear
when these correlations $r_{A_{i}B_{a}}$ are calculated via {\it different} theoretical models with  emphasis on the transport model involved.

Now, after checking
 all $r_{A_{i}A_{j}}$ (Eq.(\ref{r XY})) consistency conditions
$r_{A_{i}A_{j}} \in \langle -1, 1\rangle$, $i,j=1,2,3$, $i \neq j$, three partial correlation
coefficients
$(r_{A_{i}A_{j}|A_{k}}) = (r_{SK|L}, r_{LK|S}, r_{SL|K})$
between
force parameters $K_{0}$, $S_{0}$ and $L$ can be calculated as follows \cite{Kleinbaum,Kendall}:
\begin{eqnarray}
\label{r KS|L}
r_{A_{i}A_{j}|A_{k}} = \frac{r_{A_{i}A_{j}} - r_{A_{i}A_{k}} r_{A_{j}A_{k}}}{\sqrt{1 - r_{A_{i}A_{k}}^{2}} \sqrt{1 - r_{A_{j}A_{k}}^{2}}} \; .
%
\end{eqnarray}

Similarly,
only these pairs of the experiments $"a"$ and $"b"$, $a \neq b$, are accepted
for which the consistency conditions $r_{A_{i}A_{j}|A_{k}} \in \langle -1, 1\rangle$ are fulfilled.
Due to Eq.(\ref{r KS|L}) the conditions
$r_{A_{i}A_{j}|A_{k}} \in \langle -1, 1\rangle$, $i,j,k=1,2,3$, $i \neq j \neq k$, give three intervals for $r_{SK}$, $r_{SL}$ and $r_{KL}$, and it has to be checked if (self-consistently) the values
$r_{SK}$, $r_{SL}$ and $r_{KL}$ belong to them.

Finally, three different multiple correlation coefficients $r_{A_{i}|A_{j}A_{k}}$, $i,j,k = 1,2,3$, $i\neq j\neq k$, which characterise the liner correlation of $A_{i}$ on both $A_{j}$ and $A_{k}$ can be calculated:
\begin{eqnarray}
\label{r A|AA}
r_{A_{i}|A_{j}A_{k}} = \sqrt{1 - (1 - r_{A_{i}A_{j}}^{2})(1 - r_{A_{i}A_{k}|A_{j}}^{2})} \; .
\end{eqnarray}
The consistency condition requires the acceptance of only these cases  for which $0 \leq r_{A_{i}|A_{j}A_{k}} \leq 1$. Yet, if both complete and partial correlation coefficients fulfill the consistency conditions then from Eq.(\ref{r A|AA}) it follows that every multiple correlation coefficient also fulfils it.


\section{Results of the numerical analysis and conclusions}
\label{numerical analysis}
\subsection{Consistency analysis}
\label{Incons analysis}
In what follows the isospin observables $B_a$, $a=1,2,...,5$ are used to identify the experiment $"a"$. Taking into account all consistency conditions for both the complete and partial correlation coefficients, the numerical analysis based on Eq.(\ref{r XY}) and Eq.(\ref{r KS|L})
gives the following results:
\begin{itemize}
\item for the energy 50 MeV/u   results of the experiments: $B_1$-$B_2$, $B_1$-$B_3$, $B_1$-$B_4$, $B_1$-$B_5$; $B_2$-$B_4$  are inconsistent
\item for the energy 120 MeV/u  results of the experiments:  $B_1$-$B_4$, $B_1$-$B_5$; $B_2$-$B_4$; $B_3$-$B_5$; $B_4$-$B_5$ are inconsistent.

\end{itemize}
From this it can be inferred that the statistical inconsistency (see Section~\ref{Basic and consistency}) for
$B_1$-$B_2$ and $B_1$-$B_3$ vanishes with the increase of the energy.
The results of the experiments $B_1$-$B_4$, $B_1$-$B_5$ and $B_2$-$B_4$ are inconsistent for both energy values, and therefore, by this analysis alone, one cannot claim  if these results
have mainly
the statistical inconsistency or whether
some theoretical inconsistences are present.
\\
Now, according to \cite{covariance-analysis-of-symmetry-energy}, one should expect better accuracy of the  measurements with  the energy increase. Therefore, it is possible that the results of the experiments $B_3$-$B_5$; $B_4$-$B_5$, whose inconsistency has been detected only for 120 MeV/u,
reflect mainly
the theoretical inconsistency (see Section~\ref{Basic and consistency}).
This suggests that
the results for the experiments $B_1$-$B_5$ reveal also the theoretical inconsistency (seen for both energies).
Therefore, the above analysis suggests caution when comparing results from experiments
 that involve as the isospin sensitive observable the
diffusion of the nucleons in the neck region during the nuclear collisions quantified by the isospin transport ratios \cite{Tsang}, $B_5 \equiv R_{diff}$, with those inferred from the analysis of
the remaining $B_a$, $a=1,2,3,4$, observables.
\subsection{Correlation analysis results}
\label{Corr results}
For  energy 50 MeV/u (Figure~1) the symmetry energy coefficient $S_{0}$ depends
moderately or
weakly on the incompressibility of the pure neutron matter $K_{0}$ and the slope of the symmetry energy $L$. For example for the latter case  $r_{SL|K} \in (-0.45, 0.34)$.
\\
When the  energy increases to
120 MeV/u (Figure~2) then the linear dependance of $S_{0}$ on $L$ becomes much stronger, i.e., $r_{SL|K} \in (-0.92, -0.71)$
except the results
connected with $B_{5} \equiv R_{diff}$ that for  $B_{5}-B_{2}$ gives $r_{SL|K} = 0.17$.
The other
partial correlations are as follows:
for 50 MeV/u, $r_{SK|L} \in (-0.59, 0.37)$
and $r_{LK|S} \in (-0.22, 0.81)$.
For 120 MeV/u, $r_{LK|S} \in (0.02, 0.46)$ and $r_{SK|L} \in (0.33, 0.56)$,
with the exception of the negative value result for $B_2-B_5$, for which
 $r_{SK|L} = -0.76$. Once again there is a problem with $B_{5}$.
\begin{figure}\sidecaption
\resizebox{0.50\textwidth}{!}
{%
  \includegraphics{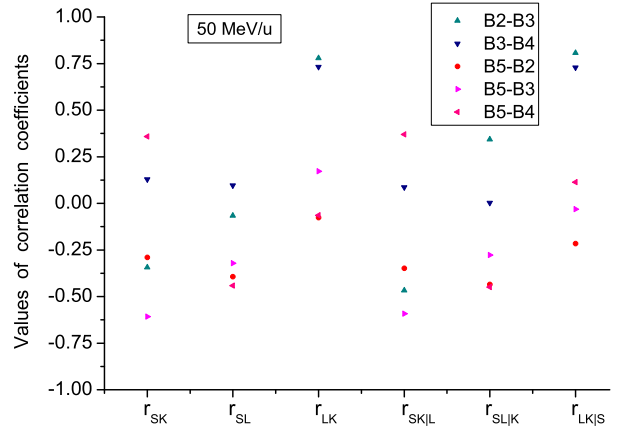}
}
\caption{Complete and partial correlation coefficients (50 MeV/u).}
\label{korSLK120}
\end{figure}
\begin{figure}
\resizebox{0.50\textwidth}{!}{%
  \includegraphics{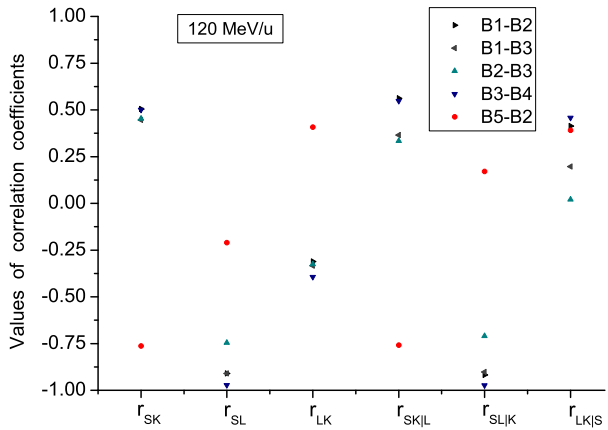}
}
\caption{
Complete and partial correlation coefficients
(120 MeV/u).}
\label{korSLK120}
\end{figure}
Having obtained
the complete and partial correlation coefficients,
 the value of the multiple correlation coefficient $r_{S|LK}$ has been obtained from
Eq.(\ref{r A|AA}). It increases
with  energy, mainly as the result of the
behaviour of $r_{SL|K}$.
For 50 MeV/u, $r_{S|LK}$ lies in the region $(0.13, 0.65)$ and for 120 MeV/u
in the region $(0.77, 0.98)$. Thus, with the increase of energy
the linear correlation of $S_{0}$ with both $L$ and $K_{0}$ together becomes stronger.
As the
scatter
of these correlation coefficients
decreases
with  energy,
the value of $r_{S|LK}$ stabilises, which most likely is connected with the increase of the accuracy of the experiments with the energy \cite{covariance-analysis-of-symmetry-energy}.
\subsection{Final conclusions}
\label{Final conclusions}
The conclusions are twofold. Firstly, it has been shown that along with the increase of energy per nucleon there are classes of observables for which the correlation coefficients for the force parameters are very close in value. This  indicates the existence of strong  $r_{S|LK}$ correlation for these classes of observables.
Secondly, an out-of-sample validity check has been performed. This  refers to the extent to which a given
result obtained in \cite{covariance-analysis-of-symmetry-energy} for a particular observable $B_{a}$   can be considered
consistent with the
 results
obtained from another isospin sensitive observable $B_{b}, a\neq b$. The performed analysis points to the existence of some inconsistences between results got with the use of different observables.
The consistency tests reveal, for the higher value of energy, groups of observables for which the calculated correlation coefficients reach very close values, with the exception of $B_{5}$ (see Figure 2).
The predicted and measured isospin sensitive observables $B_{a}, (a=1,\ldots 5)$ depend crucially on many conditions with the density dependence of the symmetry energy being the most important one \cite{2011:Coupland}.
The relevant factor is connected with the fact that these observables  probe the isospin dependent part of the EoS at different physical conditions of which the density and the value of isospin asymmetry $\delta$ play the key role. This applies in particular to the parameter $B_{5} \equiv  R_{diff}$  denoting isospin diffusion and sheds new light on the $B_{5}$ modelling problem.

\section*{Authors contributions}
All the authors were involved in the preparation of the manuscript.
All the authors have read and approved the final manuscript.
%
%

\end{document}